\documentclass[aps,nofootinbib,superscriptaddress]{revtex4}
\usepackage{epsfig}
\usepackage{url}

\newcommand{\ben}{\begin{eqnarray}}
\newcommand{\een}{\end{eqnarray}}
\newcommand{\nnu}{\nonumber\\}
\newcommand{\bef}{\begin{figure}[htb]\centering}
\newcommand{\eef}{\end{figure}}
\newcommand{\sla}[1]{{#1}\!\!\!\slash}

\begin{document}
\title{An Observation Concerning the Process Dependence of the Sivers
Functions}
\author{Zhong-Bo Kang}
\email{zkang@bnl.gov}
\affiliation{RIKEN BNL Research Center,
                  Brookhaven National Laboratory,
                  Upton, NY 11973, USA}

\author{Jian-Wei Qiu}
\email{jqiu@bnl.gov}
\affiliation{Physics Department,
                  Brookhaven National Laboratory,
                  Upton, NY 11973, USA}
\affiliation{C.N. Yang Institute for Theoretical Physics,
                  Stony Brook University,
                  Stony Brook, NY 11794, USA}
\affiliation{Department of Physics and Astronomy,
                  Iowa State University,
                  Ames, IA 50011, USA}

\author{Werner Vogelsang}
\email{werner.vogelsang@uni-tuebingen.de}
\affiliation{Institute for Theoretical Physics,
                  Universit\"{a}t T\"{u}bingen,
                  Auf der Morgenstelle 14,
                  D-72076 T\"{u}bingen, Germany}

\author{Feng Yuan}
\email{fyuan@lbl.gov}
\affiliation{Nuclear Science Division,
                   Lawrence Berkeley National Laboratory,
                   Berkeley, CA 94720, USA}
\affiliation{RIKEN BNL Research Center,
                  Brookhaven National Laboratory,
                  Upton, NY 11973, USA}

\begin{abstract}
The $k_\perp$-moment of a quark's Sivers function is known to be
related to the corresponding twist-three quark-gluon correlation 
function $T_{q,F}(x, x)$. The two functions have been extracted 
from data for single-spin asymmetries in semi-inclusive deep 
inelastic scattering and in single-inclusive hadron production 
in $pp$ collisions, respectively. Performing a consistent 
comparison of the extracted functions, we find that they show a
``sign mismatch'':
while the magnitude of the functions is roughly consistent, the 
$k_\perp$-moment of the Sivers function has opposite sign from that
of $T_{q,F}(x, x)$, both for up and for down quarks. Barring any
inconsistencies in our theoretical understanding of the Sivers
functions and their process dependence, the implication of this
mismatch is that either, the Sivers effect is not dominantly
responsible for the observed single-spin asymmetries in $pp$ 
collisions or, the current semi-inclusive lepton scattering 
data do not
sufficiently constrain the $k_\perp$-moment of the quark 
Sivers functions. Both possibilities strengthen the case for further 
experimental investigations of single-spin asymmetries in high-energy
$pp$ and $ep$ scattering. 
\end{abstract}

\date{\today}
\maketitle

\section{Introduction}
\label{sec:intro}

Since the observation of surprisingly large single transverse spin 
asymmetries (SSAs) in $p^\uparrow p\to \pi X$ at Femilab in the
1980s \cite{SSA-fixed-tgt}, the exploration of the physics behind the
observed SSAs has become a very active research branch in hadron
physics, and has played an important role in our efforts to
understand QCD and nucleon structure \cite{D'Alesio:2007jt,
Barone:2010zz}. Defined as
$A_N=\left(\sigma(s_\perp)-\sigma(-s_\perp)\right)/\left(\sigma(s_\perp)+
\sigma(-s_\perp)\right)$,
the ratio of the difference and the sum of the cross sections when
the hadron's spin vector $s_\perp$ is flipped, significant SSAs have by 
now been consistently observed in various experiments at different
collision energies. These include semi-inclusive hadron
production at low transverse momentum 
$P_{h\perp}$ in deep-inelastic scattering, $\ell
N^\uparrow\to \ell' h X$, by the HERMES Collaboration at DESY
\cite{HERMES}, COMPASS at CERN \cite{COMPASS}, and
CLAS at Jefferson Lab \cite{jlab}, as well as
inclusive single-hadron production at high $P_{h\perp}$ in
hadron-hadron collisions, $p^\uparrow p\to hX$, by the STAR, PHENIX,
and BRAHMS collaborations at RHIC \cite{SSA-rhic}. The observed
large size of SSAs in hadronic scattering initially presented a
challenge for QCD theorists \cite{Kane:1978nd}. Later two
complementary mechanisms were proposed to describe the
measured SSAs, and both of them have been quite successful
phenomenologically \cite{Collins:2005ie,Anselmino:2005ea,
Anselmino:2008sga, DY, Anselmino:2007fs, newpion, Kanazawa:2010au,
Kang:2008qh, Anselmino}.

One mechanism relies on the so-called transverse momentum
dependent (TMD) factorization
\cite{Collins:1981uk,TMD-fac,Collins:2004nx,Brodsky,MulTanBoe,boermulders,mulders,mulders1},
and describes the SSAs in terms of the spin-dependent part of TMD
parton distribution functions (PDFs), known as the Sivers
functions \cite{Siv90}, or TMD fragmentation functions (FFs),
known as the Collins functions \cite{Collins93}. This TMD
factorization approach is suitable for evaluating the SSAs of
scattering processes with two very different momentum scales,
$Q_1\gg Q_2\gtrsim \Lambda_{QCD}$. The larger scale $Q_1$ is
necessary for using perturbative QCD, while the lower scale $Q_2$
makes the observable sensitive to the parton's transverse 
motion. For example, the SSAs of hadron production at
low $P_{h\perp}$ in lepton-hadron semi-inclusive deep inelastic
scattering (SIDIS) have this characteristic property: $Q\gg
P_{h\perp}\sim \Lambda_{\rm QCD}$, and can be studied within the
TMD factorization approach.

The other mechanism generalizes the successful leading-power QCD
collinear factorization formalism to the next-to-leading power in
the expansion in $1/Q$, where $Q$ is the large momentum transfer of the
collision, and describes the SSAs in terms of twist-3
transverse-spin-dependent three-parton correlation functions
\cite{Efremov,qiu_spin,oldpion,koike}, or a combination of the
transversity distribution and three-parton fragmentation functions
\cite{oldpion,koike,finalstate}.  This so-called twist-3 collinear
factorization approach is more relevant to the SSAs for
processes in which all observed momentum transfers $Q$ are much larger
than $\Lambda_{QCD}$. This applies, for example, to the SSAs of 
inclusive single hadron production at high $P_{h\perp}$ in $p^\uparrow p$ 
collisions.  Although the two mechanisms
describe the SSAs in two very different kinematic domains, they
were shown to be equivalent in the overlap region where they both
apply, and they thus provide a unified QCD description for the
SSAs \cite{unify}.

One of the potentially important contributions to the SSAs is the Sivers
effect, which is generated by the initial- and final-state
interactions between the struck parton and the spectators or the
remnant of the polarized hadron \cite{Brodsky}.  The interactions
provide the necessary phase that leads to the non-vanishing SSAs.
In the TMD factorization approach, the role of these interactions
is accounted for by including the appropriate color gauge links
into the definition of the TMD parton distributions, whose
spin-dependent part defines the Sivers functions \cite{TMD-dis,
boermulders, Collins:2002kn}. Since the details of the initial-
and final-state interactions depend on the color flow of the
scattering process, the form of the gauge links including the
phase of the interactions is process dependent.   Since the
gauge links are included in the definition of the TMD parton
distributions, the Sivers functions, too, are found to be process
dependent \cite{mulders}.  Due to parity and time-reversal
invariance of the strong interactions, the process dependence of
the Sivers functions is effectively reduced to a sign change
between their definitions in SIDIS and in Drell-Yan lepton-pair 
production in $p^\uparrow p$ collisions \cite{boermulders, Collins:2002kn}.  
The predictive power of the TMD factorization approach relies on this
modified universality of the Sivers functions.

On the other hand, in the twist-3 collinear factorization
approach, the process dependence of the initial- and final-state
interactions is absorbed into the short-distance perturbative
hard-part functions, while keeping the relevant twist-3
three-parton correlation functions universal or
process-independent.  The necessary phase for generating the
non-vanishing SSAs arises from the quantum interference between a
scattering amplitude with one active collinear parton and an
amplitude with two active collinear partons.  The SSAs are
therefore proportional to the non-probabilistic three-parton
correlation functions.  Unlike the TMD parton distributions, which
at given transverse momentum provide direct information on a 
parton's transverse motion, the twist-3 three-parton correlation
functions provide a net asymmetry of the parton's transverse motion,
after integration over all values of the parton's transverse momentum.
As a result, the twist-3 three-parton correlation functions have a
close connection with the transverse momentum $k_\perp$-moment of
TMD parton distributions.   More precisely, the twist-3
quark-gluon correlation function, $T_{q,F}(x, x)$, often referred
to as Efremov-Teryaev-Qiu-Sterman (ETQS) function, is equal
to the first $k_\perp$-moment of the quark Sivers function
$f_{1T}^{\perp q}(x, k_\perp^2)$ probed in SIDIS (or Drell-Yan) 
processes \cite{boermulders,unify,Ma:2003ut}.

Following the tremendous progress in experimental measurements of
SSAs in recent years, the quark Sivers functions and the ETQS
functions have been extracted for various quark flavors from 
the single-spin asymmetries in SIDIS and in $pp$ scattering,
respectively.  In this paper, we examine the existing
parameterizations of these two functions to see whether the first
$k_\perp$-moments of Sivers functions are consistent with the
existing twist-3 ETQS functions.  Taking the quark Sivers functions
$f_{1T}^{\perp q}(x, k_\perp^2)$ extracted from SIDIS
\cite{Anselmino:2005ea, Anselmino:2008sga}, we evaluate their
first $k_\perp$-moments, and derive the ETQS functions $T_{q,F}(x,
x)$ with the help of the operator relation between the two functions
\cite{boermulders,unify,Ma:2003ut}.  We then compare the resulting 
``indirectly'' obtained quark-gluon correlation functions with those
``directly'' extracted from the global fit \cite{newpion} to the SSAs for
inclusive single hadron production in $p^\uparrow p$ collisions.  
In doing so, we first observe that the sign convention adopted
for the SSA in $p^\uparrow p\to hX$ in the previous 
literature \cite{qiu_spin,oldpion,newpion} is in fact not 
consistent with that used for the experimental data. As a result,
the signs of the $T_{q,F}(x,x)$ functions extracted in \cite{newpion}
need to be reversed. {\it After this adjustment, we find that the 
twist-3 correlation functions $T_{q,F}(x, x)$ obtained in the 
two different ways have conflicting signs}. 

The rest of our paper is organized as follows.  In the next
section, we briefly review the definitions of the quark Sivers
functions and the twist-3 quark-gluon correlation functions (or
ETQS functions). We recall the operator relation between the
$k_\perp$-moment of the Sivers functions and the ETQS functions, and
discuss its limitations and the corrections to it. In 
Sec.~\ref{sec:mismatch}, we present our findings regarding the 
sign ``mismatch'' between the existing parameterizations of 
quark Sivers functions and twist-3 quark-gluon correlation functions.
We discuss the possible origins of this mismatch, and potential remedies. 
We also address the implications for phenomenology and
propose further measurements to test the mechanisms for 
SSAs in hadronic processes. Finally, we give our conclusions 
and summary in Sec.~\ref{sec:summary}. An Appendix describes
the derivation of the correct signs of the ETQS functions in
single-inclusive hadron production in $pp$ scattering.

\section{The Sivers functions and the ETQS functions}
\label{sec:relation}

In this section, we recall the definitions and relations between
the quark Sivers functions and the twist-3 quark-gluon correlation
functions, or ETQS functions.  We use light-cone coordinates
with the two light-like vectors \ben
\bar{n}^\mu=[1^+,0^-, 0_\perp], \qquad n^\mu=[0^+,1^-, 0_\perp],
\een to project out the light-cone components: $v^+ = v^\mu\,
g_{\mu\nu}\, n^\nu$ and $ v^- = v^\mu\, g_{\mu\nu}\, \bar{n}^\nu$
of any four-vector $v^\mu$.  For the fully antisymmetric tensor
$\epsilon^{\mu\nu\rho\sigma}$, we adopt the convention
$\epsilon^{0123}=1$. We choose a frame in which the momentum of
the transversely polarized hadron, $p$, is in the ``$+z$''
direction, with no transverse components: $p^\mu=p^+\bar{n}^\mu$.

The quark Sivers functions for SIDIS kinematics with the
transversely polarized proton moving in the $+z$-direction is
defined through the following quark-field correlator
\cite{Bacchetta:2006tn} \ben {\cal M}(x, k_\perp)
=\int\frac{d\xi^- d^2\xi_\perp}{(2\pi)^3}\, e^{ixp^+
\xi^-}e^{-i\vec{k}_{\perp}\cdot \vec{\xi}_{\perp}} \langle
p,s_\perp|\bar{\psi}(0)W_{[0,\xi]}\psi(\xi)|p,s_\perp\rangle|_{\xi^+
= 0}, \label{mx} \een where \ben W_{[0,\xi]}={\cal
P}\exp\left[ig\int_0^\infty d\eta^- A^+(\eta^-, 0_\perp)\right]
{\cal P}\exp\left[ig\int_{0_\perp}^{\xi_\perp} d\eta_\perp
A_\perp(\infty^-, \eta_\perp)\right] {\cal
P}\exp\left[ig\int_\infty^{\xi^-} d\eta^- A^+(\eta^-,
\xi_\perp)\right] \label{link} \een is the gauge link consistent
with the SIDIS process and ${\cal P}$ indicates path ordering
\cite{TMD-dis}.  We note that the gauge link depends on the
transverse separation $\xi_\perp$ of the two field operators,
which is responsible for the process dependence of TMD parton
distribution functions \cite{Collins:2002kn}.  Here it is worth
pointing out that a different sign convention for the strong
coupling constant $g$ (for the interaction between the quark and the
gluon) would lead to a different sign in the exponent of the gauge
link in Eq.~(\ref{link}) (i.e., from $ig$ to $-ig$).  For deriving
Eq.~(\ref{link}) we adopted the convention of the covariant
derivative $D_\mu$ as \ben D_\mu=\partial_\mu+igA_\mu,
\label{gsign} \een for the relevant part of the QCD Lagrangian density
${\mathcal L}=\bar{\psi}i\gamma^\mu D_\mu\psi$.  Different
conventions for $D_\mu$ (such as $D_\mu=\partial_\mu-ig A_\mu$)
exist in the literature and in textbooks. Different conventions
usually do not introduce any difference for a cross section, which
has an even power in $g$.  However, the single transverse spin
asymmetry is proportional to the difference of two cross sections
with the spin flipped, and the asymmetry is a consequence of an
interference between scattering amplitudes of different phases,
which is linearly proportional to $ig$.  Therefore, one has to use
the convention consistently in the theoretical definition and
calculation of the single transverse spin asymmetry.

Following the so-called Trento convention \cite{Bacchetta:2004jz}, the
correlator ${\cal M}(x, k_\perp)$ can be expanded as
\ben {\cal M}(x, k_\perp)=\frac{1}{2}\left[f_1(x,
k_\perp^2)\sla{\bar{n}}+ \frac{1}{M}f_{1T}^\perp(x,
k_\perp^2)\epsilon^{\mu\nu\rho\sigma}\gamma_\mu \bar{n}_\nu
k_{\perp \rho} s_{\perp\sigma}\right], \label{trento} \een where
$M$ is the nucleon mass, $f_1(x, k_\perp^2)$ is the spin-averaged
TMD PDF, and $f_{1T}^\perp(x, k_\perp^2)$ is the quark Sivers
function. We note that a different convention for the Sivers 
function is commonly adopted in the phenomenological studies by the
Torino group \cite{Anselmino:2005ea,Anselmino:2008sga}. Here
a function $\Delta^N f_{q/A^\uparrow}(x, k_\perp)$  is introduced
which is defined from \ben f_{q/A^\uparrow}(x,
k_\perp)\equiv {\rm Tr}\left[\frac{1}{2} \sla{n} \, {\cal M}(x,
k_\perp)\right] = f_1(x, k_\perp^2)+\frac{1}{2}\Delta^N
f_{q/A^\uparrow}(x, k_\perp)\,{\mathbf s}_\perp\cdot (\hat{\mathbf
p}\times \hat{\mathbf k}_\perp). \een 
The relation between $\Delta^N f_{q/A^\uparrow}$ and the Sivers
function in the Trento
convention is \ben \Delta^N f_{q/A^\uparrow}(x,
k_\perp)=-\frac{2k_\perp}{M} f_{1T}^{\perp q}(x, k_\perp^2).
\label{conm} \een

In the twist-3 collinear factorization approach, the ETQS
function $T_{q, F}(x, x)$
is defined as \cite{newpion,
Kang:2008qh} \ben T_{q,F}(x, x)=\int\frac{d\xi^-
d\zeta^-}{4\pi}e^{ixp^+\xi^-} \langle p,s|\bar{\psi}(0)V_{[0,
\zeta]}\gamma^+\left[ \epsilon^{s_\perp\sigma n\bar{n}}F_\sigma^{~
+}(\zeta^-)\right] V_{[\zeta, \xi]} \psi(\xi^-)|p,s\rangle,
\label{tqf} \een where $V_{[0, \zeta]}$ and $V_{[\zeta, \xi]}$ are
the gauge links along the ``$-$'' light-cone direction 
and are given by \ben
V_{[\zeta, \xi]}={\cal P} {\rm
exp}\left[ig\int_{\zeta^-}^{\xi^-}d\eta^- A^+(\eta^-)\right].
\label{colllink} \een Here the sign convention for coupling
constant $g$ is the same as that in Eq.~(\ref{gsign}). Choice
of, for example, the convention with $D_\mu=\partial_\mu-ig A_\mu$ 
would change the sign of the exponent. Within the
collinear factorization approach, it is assumed that the typical
transverse momentum of all active partons is much smaller than the
hard scale of the scattering process, $Q$.  Up to power corrections
in $1/Q$, the transverse momenta of all active partons are completely
integrated into non-perturbative PDFs, FFs, or the correlation
functions.  Consequently, unlike for the TMD distributions, all field
operators defining the non-perturbative functions in the collinear
factorization approach are evaluated at the same light-cone separation 
with zero ``$+$'' and ``$\perp$'' components, as shown for example in
Eq.~(\ref{tqf}).

Since the quark-gluon correlation functions in the collinear
factorization approach have all their active partons' transverse
momenta integrated, these correlation functions can be related
to $k_\perp$-moments of the TMD parton distribution functions.  It
was shown at the operator level \cite{boermulders,unify,Ma:2003ut}
that the ETQS function $T_{q,F}(x, x)$ is closely related to the
$k_\perp$-moment of Sivers function: \ben gT_{q, F}(x, x)=-\int
d^2k_\perp \frac{|k_\perp|^2}{M}f_{1T}^{\perp q}(x,
k_\perp^2)|_{\rm SIDIS} \label{rel} \een where the subscript
``SIDIS'' emphasizes that the Sivers functions here are probed in
the SIDIS process. We stress again the importance of the
sign convention for the coupling constant $g$ in the definition of
the gauge link. If the sign convention used to define $T_{q,F}(x,
x)$ is different from that in the definition of $f_{1T}^{\perp
q}(x, k_\perp^2)$, the difference will
introduce an extra factor ``$-1$'' in the relation between these
two functions, so that there will be no minus sign on the
right-hand side of Eq.~(\ref{rel}).

We emphasize that the operator definition
in Eq.~(\ref{tqf}) does not completely fix the quark-gluon correlation
function $T_{q, F}(x, x)$, unless the renormalization
scheme is specified. As is well known from the case
of ordinary PDFs, the matrix 
element in Eq.~(\ref{tqf}) is ultraviolet (UV) 
divergent \cite{Collins:2003fm}. Like in the case of PDFs, 
the quark-gluon correlation function is really
defined in terms of the QCD factorization formalism.  The leading UV
divergent (the large $k_\perp$) region of the matrix element on
the right-hand-side of Eq.~(\ref{tqf}) corresponds to the region of
phase space with large parton virtuality, and is required by
factorization to be moved from the matrix element into the
perturbatively calculated short-distance functions.
The removal or subtraction of the UV divergence is not unique,
which leads to the factorization scheme and scale ($\mu$) dependence
of the correlation functions $T_{q, F}(x, x, \mu)$~\cite{evolution}.
In this way, also the relation in Eq.~(\ref{rel}) is subject to the
UV subtractions and the adopted factorization scheme, and 
hence not a unique identity.
That said, the relation~(\ref{rel}) provides a natural ``zeroth-order''
connection between the Sivers and the ETQS functions. It plays 
an important role in establishing the consistency between the TMD 
factorization approach and the collinear twist-three quark-gluon 
correlation approach in the descriptions of the SSAs in SIDIS 
and the Drell-Yan process~\cite{unify}. It also is a useful 
starting point for phenomenological studies and 
is of much help in testing the various constraints on the
quark Sivers and quark-gluon correlation functions. In the
following, we will therefore make use of relation~(\ref{rel}), keeping
however in mind the caveats we have made regarding UV renormalization.

\section{The ``sign mismatch"}
\label{sec:mismatch}

The quark Sivers functions $f_{1T}^{\perp q}(x, k_\perp^2)$ (or
equivalently, $\Delta^N f_{q/A^\uparrow}(x, k_\perp)$) and the
twist-3 quark-gluon correlation functions $T_{q, F}(x, x)$ have
been extracted from experimental data on SSAs for single hadron
production in SIDIS and in hadron-hadron scattering, respectively.
In this section, we compare the existing parameterizations of
these two functions and present our findings concerning the ``sign 
mismatch''.  We also introduce and discuss various loopholes that
might resolve the apparent inconsistency. 

So far the quark Sivers functions have been extracted from the
$A_{UT}^{\sin(\phi_h-\phi_s)}$ azimuthal asymmetries in SIDIS. 
We consider two such parametrizations here. One is from
Ref.~\cite{Anselmino:2005ea} (we refer it as ``old Sivers''), the
other one (``new Sivers'') from Ref.~\cite{Anselmino:2008sga} . 
They both parametrize the spin-averaged TMD PDFs
$f_1^q(x, k_\perp^2)$ and Sivers functions $\Delta^N
f_{q/h^\uparrow}(x,k_\perp)$ for each quark flavor $q$ in the form
\ben f_1^q(x,
k_\perp^2)&=&f_1^q(x) g(k_\perp), \label{unpol_tmd}
\\
\Delta^N f_{q/h^\uparrow}(x,k_\perp) &=&2{\cal
N}_q(x)f_1^q(x)h(k_\perp)g(k_\perp), \label{sivers_tmd}
\een
where $f_1^q(x)$ is the quark's spin-averaged collinear PDF, ${\cal
N}_q(x)$ is a fitted function whose functional form is not
relevant for our discussion below, and $g(k_\perp)$ is assumed to
have a Gaussian form, \ben g(k_\perp)=\frac{1}{\pi \langle
k_\perp^2\rangle}e^{-k_\perp^2/\langle k_\perp^2\rangle} \een with
a fitting parameter $\langle k_\perp^2\rangle$ for the width. However,
the two parameterizations adopt different functional forms for
the $k_\perp$-dependence of the Sivers function: \ben \mbox{old
Sivers:} && h(k_\perp)=\frac{2k_\perp M_0}{k_\perp^2+M_0^2},
\\
\mbox{new Sivers:} && h(k_\perp)=\sqrt{2e}\, \frac{k_\perp}{M_1}e^{-k_\perp^2/M_1^2},
\een
where $M_0$ and $M_1$ are fitted parameters.

Since for both parameterizations the $k_\perp$-dependence is
assumed to be decoupled from the $x$-dependence, we can derive 
the $x$-dependence of the associated twist-3 quark-gluon correlation 
$T_{q,F}(x,x)$ analytically, using the relation in Eq.~(\ref{rel}).
By substituting the parameterization of
the Sivers function in Eq.~(\ref{sivers_tmd}) into the
right-hand-side of Eq.~(\ref{rel}), and using the fitting
parameters extracted in Refs.~\cite{Anselmino:2005ea} and
\cite{Anselmino:2008sga}, we obtain the following 
two parameterizations for the
correlation function $T_{q,F}(x,x)$: \ben gT_{q, F}(x, x)|_{\rm
old~Sivers}&=&0.40  f_1^q(x) {\cal N}_q(x)|_{\rm old}, \label{old}
\\
gT_{q, F}(x, x)|_{\rm new~Sivers}&=&0.33  f_1^q(x) {\cal
N}_q(x)|_{\rm new}. \label{new} \een From the existing data, the
best constrained Sivers functions are those of $u$ and $d$ quarks.
Using the fitted functions ${\cal N}_q(x)|_{\rm old}$ and
${\cal N}_q(x)|_{\rm new}$ from Refs.~\cite{Anselmino:2005ea} and
\cite{Anselmino:2008sga}, respectively, we plot the drived
quark-gluon correlation functions $x \,gT_{u, F}(x, x)$ (left) and
$x\, gT_{d, F}(x, x)$ (right) in Fig.~\ref{fig:sivers}.
The dashed lines are for the quark-gluon correlation functions
obtained by using the new Sivers parameterization, while the
dotted lines are for the old Sivers parameterization.  We find
that for these ``indirectly'' obtained quark-gluon correlation
functions, $T_{u, F}(x, x)$ is positive, while $T_{d, F}(x, x)$ is 
negative. \bef \vspace*{5mm}\psfig{file=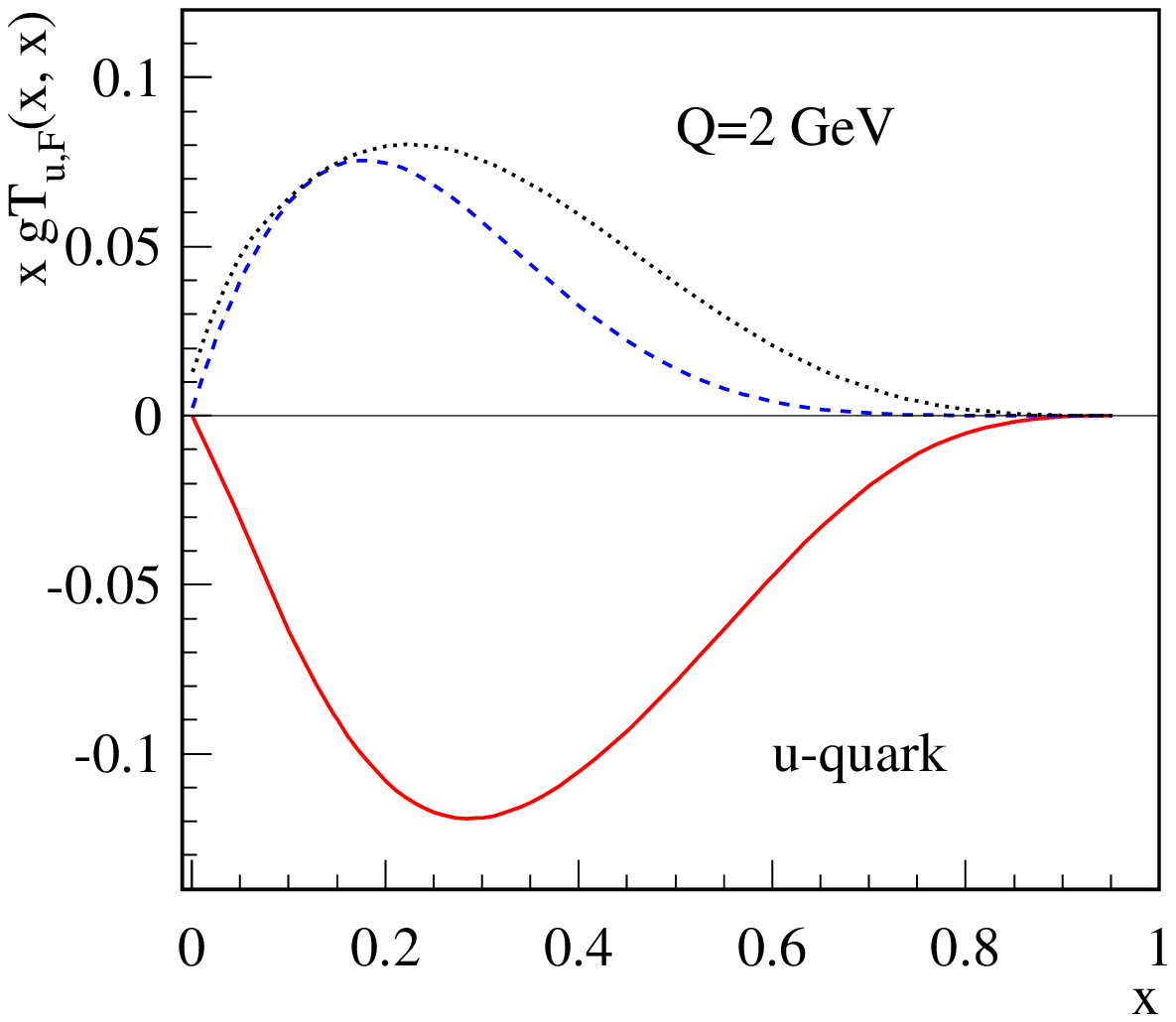, width=2.6in} \hskip
0.6in \psfig{file=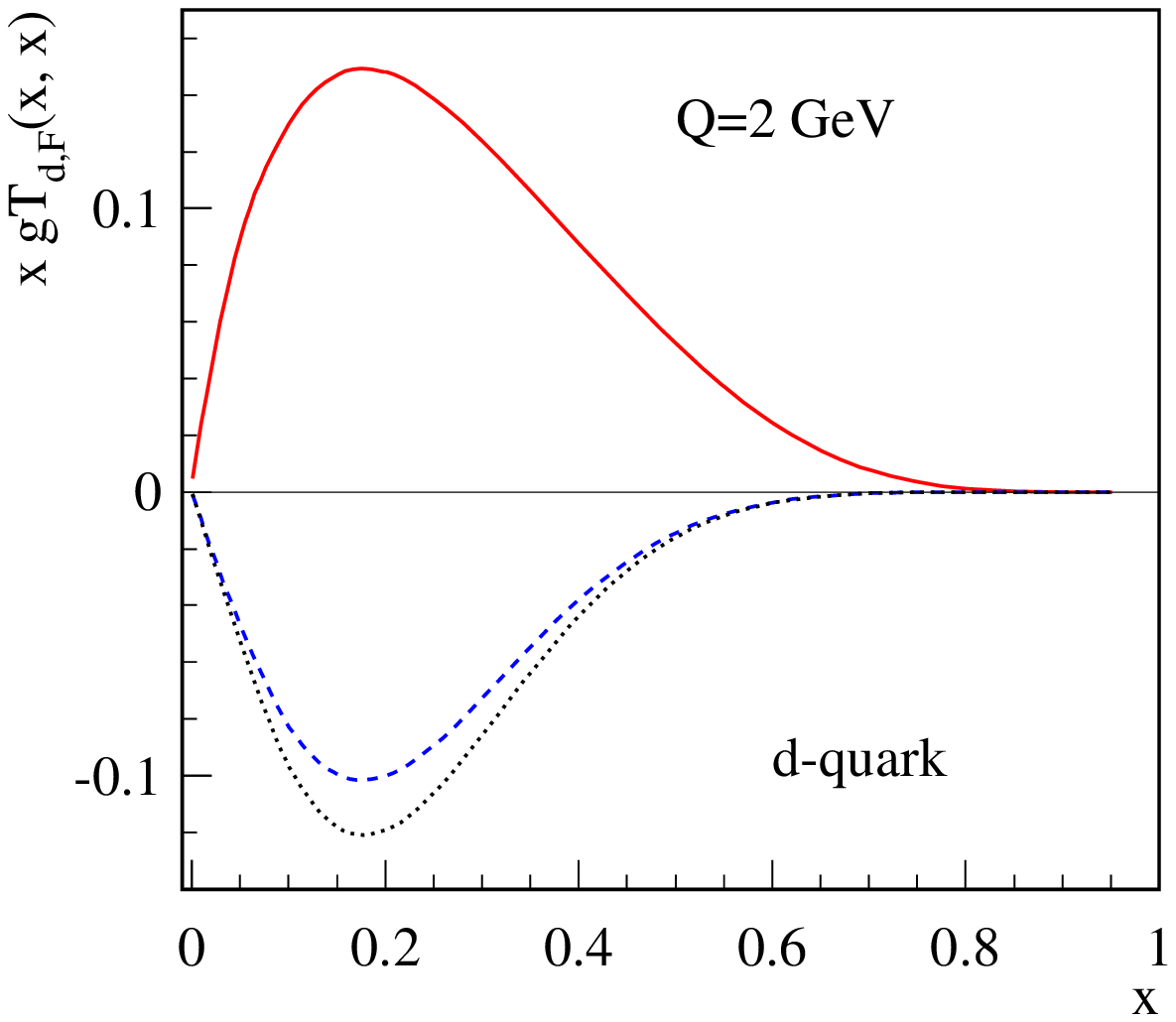, width=2.6in} \caption{The
quark-gluon correlation function $gT_{q,F}(x, x)$ as a function of
momentum fraction $x$ for $u$-quarks (left) and $d$-quarks (right).
The dashed (dotted) lines are $gT_{q, F}(x, x)|_{\rm new~Sivers}$
($gT_{q, F}(x, x)|_{\rm old~Sivers}$) obtained by taking the
$k_\perp$-moments of the corresponding quark Sivers functions 
according to the right-hand-side
of Eq.~(\ref{rel}). The solid lines represent the correlation
functions extracted directly from data on SSAs for inclusive pion
production in proton-proton collisions, $p^\uparrow p\to \pi+X$
\cite{newpion}, after correcting for the sign convention (see text).} 
\label{fig:sivers} \eef

On the other hand, the ETQS function $T_{q, F}(x, x)$ can be
``directly'' extracted from data on SSAs for inclusive single hadron
production in hadronic collisions, $p^\uparrow p\to h(P_{h\perp},y)+X$, 
assuming these asymmetries are predominantly generated by the 
Sivers effect (or rather, its twist-3 counterpart). Such SSAs have been 
measured at sufficiently large transverse momentum $P_{h\perp}$ by 
the E704 Collaboration at Fermilab \cite{SSA-fixed-tgt}, and the STAR, 
PHENIX, and BRAHMS collaborations at RHIC \cite{SSA-rhic}.
Since they depend only on one large momentum scale $P_{h\perp}$, these
SSAs are better studied in the collinear factorization approach,
where they may be generated by three possible mechanisms: (1) the
twist-3 quark-gluon and tri-gluon correlation functions of the
polarized hadron, (2) the transversity distribution of the
polarized hadron combined with the twist-3 quark-gluon
fragmentation functions to the observed hadron, and (3) the
transversity distribution combined with possible twist-3
unpolarized quark-gluon correlation functions \cite{oldpion}.
It was found that the third mechanism only makes a small 
contribution \cite{koike}.  By assuming that the observed SSAs 
are mainly generated by the ETQS functions $T_{q, F}(x, x)$, a set 
of $T_{q, F}(x, x)$ was extracted by a global fitting 
procedure \cite{newpion}. In the course of our investigations,
we have revisited the sign convention adopted in \cite{newpion},
which was based on the earlier work \cite{qiu_spin,oldpion,newpion}.
We have discovered that the convention was at odds with that
chosen in the experimental studies. The inconsistency can be traced
to the value of the contracted Levi-Civita tensor appropriate for
the spin asymmetry. We provide a detailed discussion of this
issue in the Appendix. Correcting the sign convention
of \cite{qiu_spin,oldpion,newpion} means that one needs to change 
the signs of the $T_{q, F}(x, x)$ functions extracted in \cite{newpion}.
We plot the resulting ``directly'' extracted $T_{q, F}(x, x)$ 
functions as solid lines in Fig.~\ref{fig:sivers}, along with 
the previous ones derived ``indirectly'' from the $k_\perp$-moment of the 
quark Sivers functions. Surprisingly, we find that the two sets
of functions have {\it opposite signs}, both for up and for down 
quarks.

At first sight, it may seem that we have created a problem where
none used to be. After all, the sign mismatch we find becomes
apparent only after we have changed the signs of the $T_{q, F}(x, x)$ 
functions of \cite{newpion}. However, the basic problem is easy
to see: as we discussed in the Introduction, the Sivers 
contributions to the single spin asymmetries depend on initial or 
final state interactions in the scattering processes. 
The SSA in SIDIS comes from a final state interaction.
A negative up-quark Sivers function is known to generate a positive 
SIDIS spin asymmetry for $\pi^+$ production. In $p^\uparrow p\to \pi^+X$ at 
forward rapidities, however, the main partonic channel is $ug\to ug$,
for which initial-state interactions play the dominant role, 
resulting in negative partonic hard-scattering functions. Therefore,
if the Sivers mechanism (or its twist-3 variant via Eq.~(\ref{rel})) is 
primarily responsible for the SSA in this process, one would expect a 
negative asymmetry for $\pi^+$, contrary to what is observed. Thus
the $T_{q, F}(x, x)$ functions needed to describe the RHIC single-spin
asymmetries cannot have the signs suggested by Eq.~(\ref{rel}). 
We note that in these considerations, one has to carefully take
into account the experimental definitions of the SSAs; see 
the Appendix for some details. 

There are two main caveats regarding the sign mismatch. The first 
one is that the integral over $k_\perp$ in Eq.~(\ref{rel}) might 
produce a different sign from that of $f_{1T}^{\perp q}(x, k_\perp^2)$ 
itself in the region of $k_\perp$ where it is constrained by data. 
The HERMES SIDIS data that are mostly relevant for the extraction 
of $f_{1T}^{\perp q}(x, k_\perp^2)$ are at a relatively modest $Q^2\sim 
2.4$ GeV$^2$. Since the TMD factorization formalism is valid only for 
$k_\perp \ll Q$, the data constrain the function and its sign only at 
very low $k_\perp\sim \Lambda_{\rm QCD}$. The existing parameterizations 
of the quark Sivers functions \cite{Anselmino:2005ea, Anselmino:2008sga} 
assume a purely Gaussian form of the $k_\perp$-dependence and hence would 
not allow a sign change of the function at some $k_\perp$. This leads to 
significant uncertainties in the determination of the twist-3
quark-gluon correlation functions via Eq.~(\ref{rel}), because taking 
the $k_\perp$ moment enhances the contribution from the unknown 
larger-$k_\perp$ region. Also, the issue of UV renormalization 
discussed in the previous section becomes relevant for the $k_\perp$ 
moment. We note that future SIDIS experiments at an Electron Ion Collider
would have the kinematic reach to precisely map out the $k_\perp$
dependence, and to allow measurements of the transverse-momentum 
weighted asymmetries, providing direct access to the twist-3 quark-gluon 
correlation functions. In this way, reliable comparisons with the 
correlation functions extracted from $pp$ collisions would become 
possible.

It is worth keeping in mind that the SIDIS and $pp$ single-spin asymmetry data
also probe slightly different values of $x$. The former reach up to $x\sim 0.4$,
while the latter mostly access yet larger values, $x\sim 0.6$. While it is in principle
possible that a rapid sign change could occur towards large $x$ which would
explain the mismatch, there is nothing
in the SIDIS data or the $pp$ data with a sufficiently large $x_F$ coverage that would indicate such a behavior, and we do not consider this
to be a likely scenario.

The second possibility is that there are other significant contributions
to the SSAs for single hadron production in $p^\uparrow p$ collisions, 
besides the Sivers mechanism. In addition to the asymmetry due to 
the spin-dependent twist-3 quark-gluon correlation functions of
the polarized hadron, the SSAs in hadronic collisions may also
be generated at the hadronization stage by a combination of the
transversity distribution of the polarized hadron and the twist-3
quark-gluon fragmentation functions~\cite{finalstate}, which is
effectively a representation of the Collins effect in the
collinear factorization approach. If this mechanism makes
a large contribution to the hadronic SSAs, with sign opposite
to that by the $T_{q, F}(x, x)$, it might explain the observed 
features. Unlike the measurement of SSAs in SIDIS, where the Sivers 
effect and the Collins effect can be separated by using different
azimuthal angle weighting, the two effects cannot be separated
in single-hadron inclusive production in hadronic collisions.
Nonetheless, other measurements are available in $pp$ scattering
that would allow to disentangle them. The prime example is 
the Drell-Yan process, which allows direct access to the 
Sivers or $T_{q, F}(x, x)$ functions \cite{boermulders, Collins:2002kn}.
A similar role could be played by photon pair production~\cite{marc}.
Here we will briefly consider two further processes that have the
advantage of being somewhat more copious at RHIC.

In order to get clean access to the quark-gluon 
correlation functions $T_{q, F}(x, x)$, we need observables that 
are not sensitive to the details of the hadronization stage. 
At RHIC, for example, direct photon production~\cite{mulders1} 
at large transverse momentum $P_{h\perp}$ and inclusive single jet 
production at large transverse jet energy are two promising 
observables of this kind.  Since the fragmentation contribution 
to prompt photon production at large $p_T$ is much smaller than the  
direct contribution at RHIC energies, in particular if photon 
isolation cuts are imposed, the SSAs of these two observables
could provide direct information on the twist-3 quark-gluon
correlation functions, thus allowing to see if their signs
are consistent with those derived from Eq.~(\ref{rel}). 

In Fig.~\ref{fig:prediction} we present our estimates for the SSAs 
for direct photon (left) and inclusive single jet production 
(right) in $p^\uparrow p$ collisions at $\sqrt{S}=200$~GeV. 
We consider production in the forward region of the polarized 
proton, so that to a good approximation we only need to include  
the valence quark contribution for the polarized beam. 
For the relevant unpolarized PDFs, we use those specified in 
Ref.~\cite{Anselmino:2005ea, Anselmino:2008sga, newpion} correspondingly.
The solid curves represent the SSAs calculated by
using the {\it directly} extracted $T_{q, F}(x, x)$ (with $T_{u,
F}(x, x)<0$ and $T_{d, F}(x, x)>0$), which were shown as the solid 
lines in Fig.~\ref{fig:sivers}.  The dashed and dotted curves
show the SSAs calculated by using the {\it indirectly}
derived $T_{q, F}(x, x)$ from Eq.~(\ref{old}) and Eq.~(\ref{new}),
respectively, which again were shown by the same line patterns 
in Fig.~\ref{fig:sivers} and have $T_{u,F}(x, x)>0$ and 
$T_{d, F}(x, x)<0$. The results in Fig.~\ref{fig:prediction}
demonstrate that positive $A_N$ for
direct photon and inclusive single jet production should be
expected at RHIC if the ``directly'' extracted $T_{q, F}(x, x)$
are correct. If, on the other hand, the signs of $T_{q, F}(x, x)$ 
follow Eqs.~(\ref{old}) and (\ref{new}), negative values for
the $A_N$ for the two processes are predicted. We note that 
direct photon and inclusive single jet production 
both receive contributions from the $u$ and $d$ quark ETQS functions. 
Since these have opposite signs and rather similar magnitude, 
their effects cancel to some degree for jet production. For
photons, the situation is more favorable thanks to the weighting
by the quark's charge squared, which explains why here the 
spin asymmetries are overall larger.

\bef \vspace*{5mm}
\psfig{file=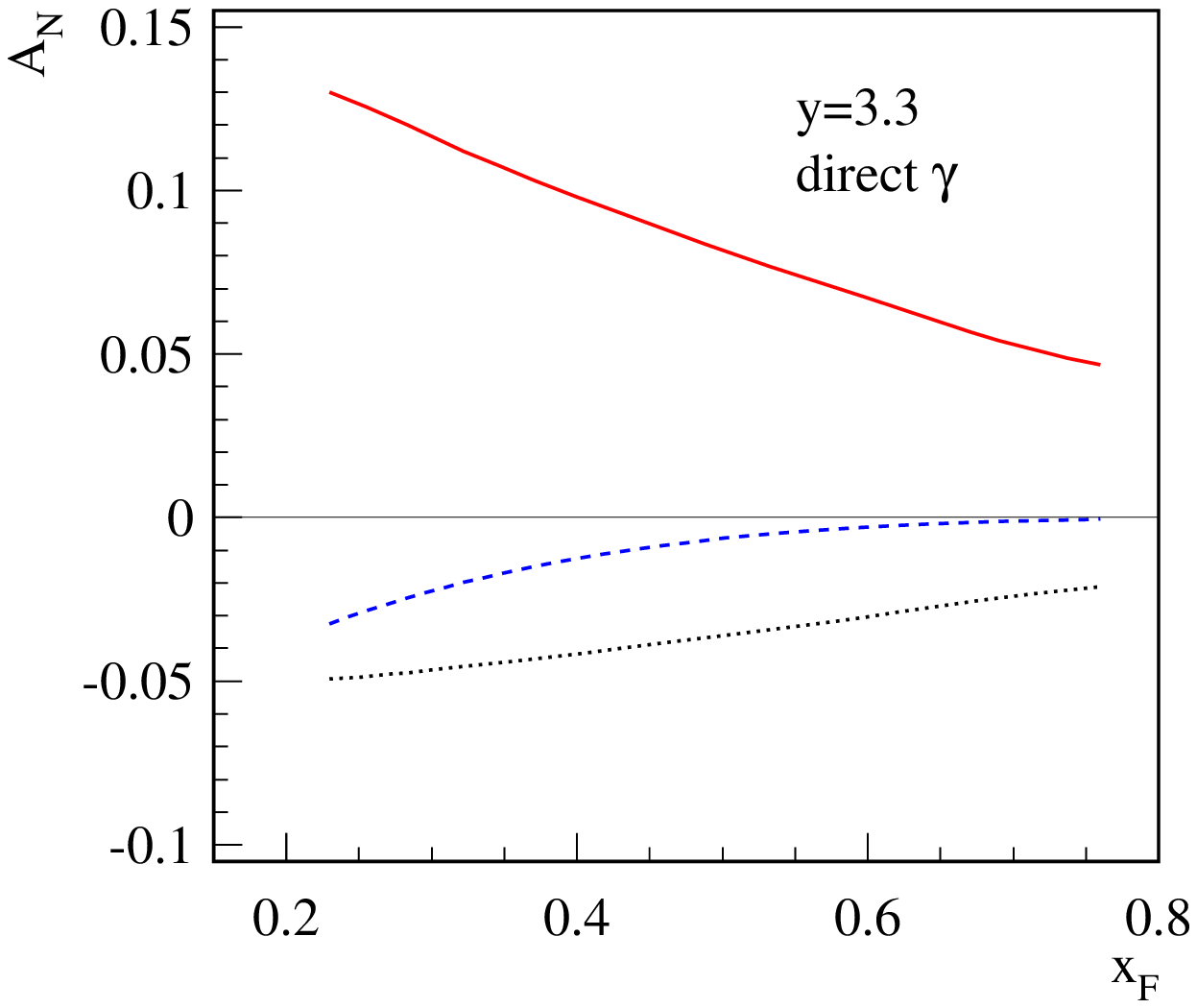, width=2.6in} \hskip 0.3in
\psfig{file=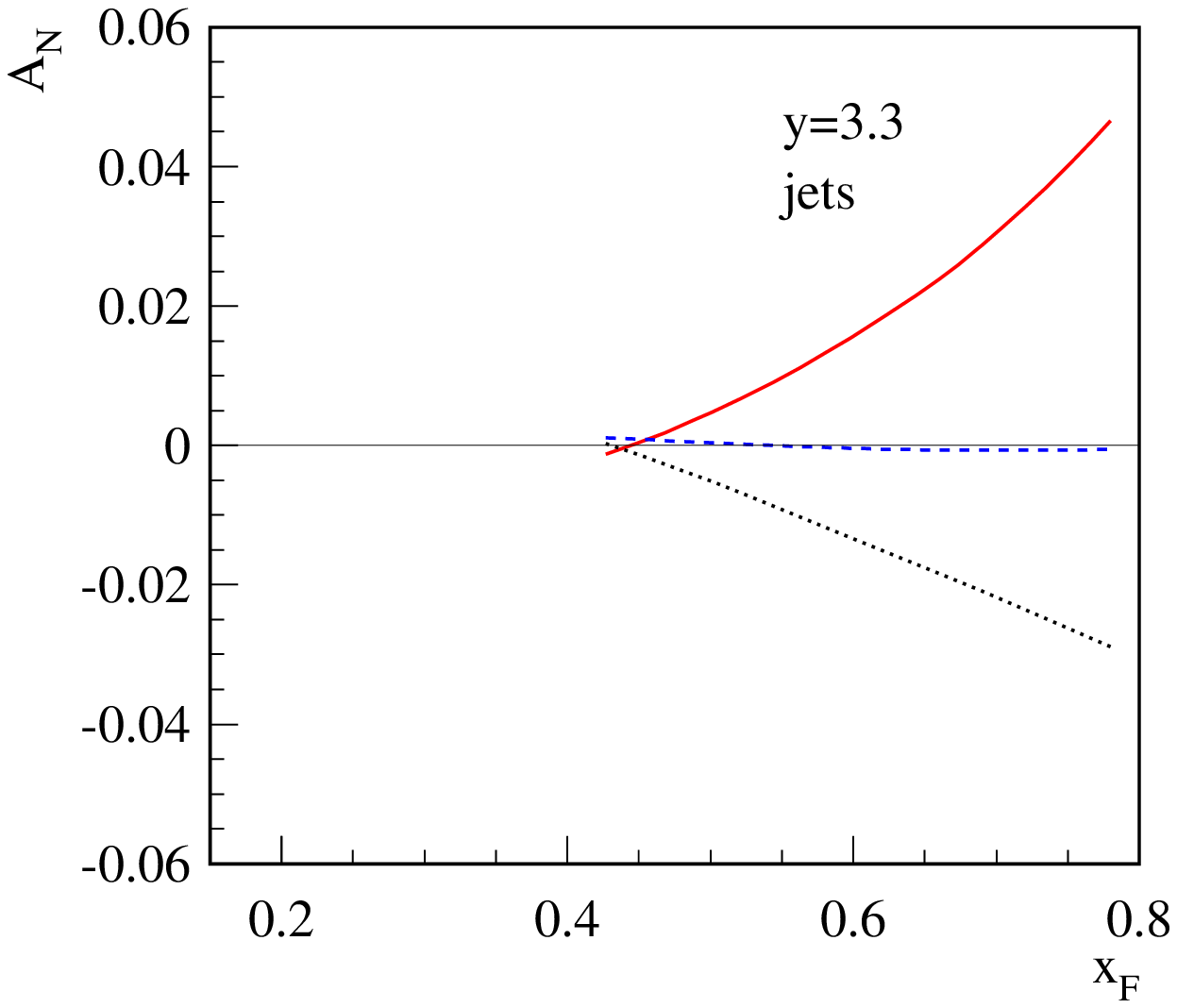, width=2.6in} \caption{The SSAs for direct
photon (left) and single inclusive jet (right) production in
$p^\uparrow p$ collisions at $\sqrt{S}=200$ GeV, as functions of 
$x_F$ for rapidity $y=3.3$. The various curves correspond to the
$T_{q, F}(x, x)$ shown in Fig.~\ref{fig:sivers}.} \label{fig:prediction} \eef

\section{Summary}
\label{sec:summary}

We have computed the $k_\perp$-moments for two parameterizations
of up and down quark Sivers functions determined from semi-inclusive 
lepton scattering data given in \cite{Anselmino:2005ea, Anselmino:2008sga}. 
These are related to the quark-gluon correlation functions $T_{q, F}(x, x)$ 
relevant for the description of single-spin asymmetries in single hadron 
production in $pp$ scattering. The latter have in the past been extracted 
from RHIC data \cite{newpion}. Correcting an inconsistency in previous 
theoretical treatments of the spin asymmetries in $pp$ scattering, we 
have found that the resulting $T_{q, F}(x, x)$ functions have signs 
opposite to those predicted from the analysis of the $k_\perp$-moments of
the Sivers functions. We have discussed various possible explanations for
this apparent discrepancy.

Our finding highlights the importance of additional measurements 
of single-spin asymmetries. Measurements of the $k_\perp$ dependence
of the Sivers functions with wide kinematic reach would be feasible
at an Electron Ion Collider and should shed light on the contributions
from various $k_\perp$-regions to the moment of the Sivers functions.
We have also shown that $A_N$ measurements for jet and direct photon 
production in $pp$ collisions at RHIC should be valuable tools for
a cleaner determination of the quark-gluon correlation functions 
$T_{q, F}(x, x)$.

\section*{Acknowledgments}
We thank H.~Avakian, L.~Gamberg, A.~Metz, B.~Musch, and A.~Prokudin for discussions and comments. 
This work was supported in part by the U.S. Department of Energy
under grant number DE-FG02-87ER4037 (JQ) and DE-AC02-05CH11231 (FY).
We are grateful to RIKEN, Brookhaven National Laboratory,
and the U.S. Department of Energy (Contract No.~DE-AC02-98CH10886)
for supporting this work.

\appendix*
\setcounter{figure}{0}
\renewcommand\thefigure{\Alph{section}.\arabic{figure}}
\section{The sign of $T_{q, F}(x, x)$ in inclusive hadron production}
\label{sec:appendix}

In this appendix, we demonstrate why the SSA data for 
$p^\uparrow p\to h X$ require $T_{u, F}(x, x)<0$ and $T_{d,
F}(x, x)>0$, if the ETQS functions are the dominant sources of 
the observed asymmetries.

We start with the QCD factorization formalism for the
spin-averaged cross section for inclusive single particle
production in hadronic collisions, $A^\uparrow(S_\perp)+B\to
h(P_{h\perp})+X$: \ben E_h\frac{d\sigma}{d^3 P_h}&=&
\frac{\alpha_s^2}{S}\sum_{a,b,c} \int \frac{dz}{z^2} D_{c\to h}(z)
\int \frac{dx'}{x'}f_{b/B}(x')\int \frac{dx}{x} f_{a/A}(x)
H^U_{ab\to c}(\hat s,\hat t,\hat u)\delta\left(\hat s+\hat t+\hat
u\right), \een where $f_{a/A}(x)$ and $f_{b/B}(x')$ are the PDFs,
$D_{c\to h}(z)$ are the FFs, and $H^U_{ab\to c}$ are the partonic 
hard-scattering functions, with $\hat s$, $\hat t$, and
$\hat u$ the Mandelstam variables at the parton level. 
Including only the contributions by the twist-3
quark-gluon correlation functions, the spin-dependent cross section
$d\Delta\sigma(s_\perp)\equiv
[d\sigma(s_\perp)-d\sigma(-s_\perp)]/2$ is given by \ben
E_h\frac{d\Delta\sigma(s_\perp)}{d^3 P_h}&=&
\frac{\alpha_s^2}{S}\sum_{a,b,c} \int \frac{dz}{z^2} D_{c\to h}(z)
\int \frac{dx'}{x'}f_{b/B}(x')\int \frac{dx}{x}
\sqrt{4\pi\alpha_s}\left(\frac{\epsilon^{P_{h\perp}s_\perp n
\bar{n}}}{z \hat{u}}\right) \nnu && \times \left[T_{a,F}(x, x) -
x\frac{d}{dx}T_{a,F}(x, x)\right] H_{ab\to c}(\hat s,\hat t,\hat
u)\delta\left(\hat s+\hat t+\hat u\right), \label{aftera} \een
where the relevant hard-scattering functions 
$H_{ab\to c}(\hat s,\hat t,\hat u)$
can be written as \ben H_{ab\to c}(\hat s,\hat t,\hat
u)=H^I_{ab\to c}(\hat s,\hat t,\hat u)+H^F_{ab\to c}(\hat s,\hat
t,\hat u)\left(1+\frac{\hat u}{\hat t}\right), \een with
$H^I_{ab\to c}$ and $H^F_{ab\to c}$ representing the contributions
from initial- and final-state interactions, respectively. The
explicit forms of $H^U_{ab\to c}$, $H^I_{ab\to c}$, and
$H^F_{ab\to c}$ are given in \cite{newpion}. It is important to
point out that the spin-dependent cross section in
Eq.~(\ref{aftera}) is calculated from an interference between two
partonic amplitudes. It thus depends on the sign convention for the
coupling constant $g$; the form given in Eq.~(\ref{aftera}) is
based on the convention in Eq.~(\ref{gsign}). If one uses the
other sign convention for the covariant derivative, there will be
an extra minus sign appearing on the right-hand side of Eq.~(\ref{aftera}),
which would be compensated by an extra sign in Eq.~(\ref{rel}). 

The SSA, $A_N$, is given by the ratio of spin-dependent and
spin-averaged cross sections:
\ben
\left.E_h\frac{d\Delta\sigma(s_\perp)}{d^3
P_h}\right/E_h\frac{d\sigma}{d^3 P_h}\equiv
A_N\sin(\phi_s-\phi_h), \label{andef} \een 
where $\phi_h$ and
$\phi_s$ are the azimuthal angles of the hadron transverse
momentum $P_{h\perp}$ and the spin vector $s_\perp$, respectively.
The absolute sign of $A_N$ depends on the choice of frame and the
coordinate system. In experiment the following convention is
used: positive values of $A_N$ correspond to a larger cross
section for hadron production to the beam's {\it left} when the
beam's proton spin is vertically {\it upward} \cite{oldpion}, as
sketched in Fig.~\ref{fig:exp}.  In the center-of-mass frame of
$A$ and $B$, a convenient coordinate system (consistent with the 
experimental convention) is given by choosing the polarized nucleon 
$A$ to move along $+z$, the unpolarized $B$ along $-z$, 
the spin vector $s_\perp$ along $y$, and the produced hadron's
transverse momentum $P_{h\perp}$ along the $x$-direction.
In this frame, $\phi_h=0$, $\phi_s=\pi/2$, and \ben
\epsilon^{P_{h\perp}s_\perp n \bar{n}}=-|P_{h\perp}||s_\perp|.
\label{epsilon} \een \bef \vspace*{3mm} 
\psfig{file=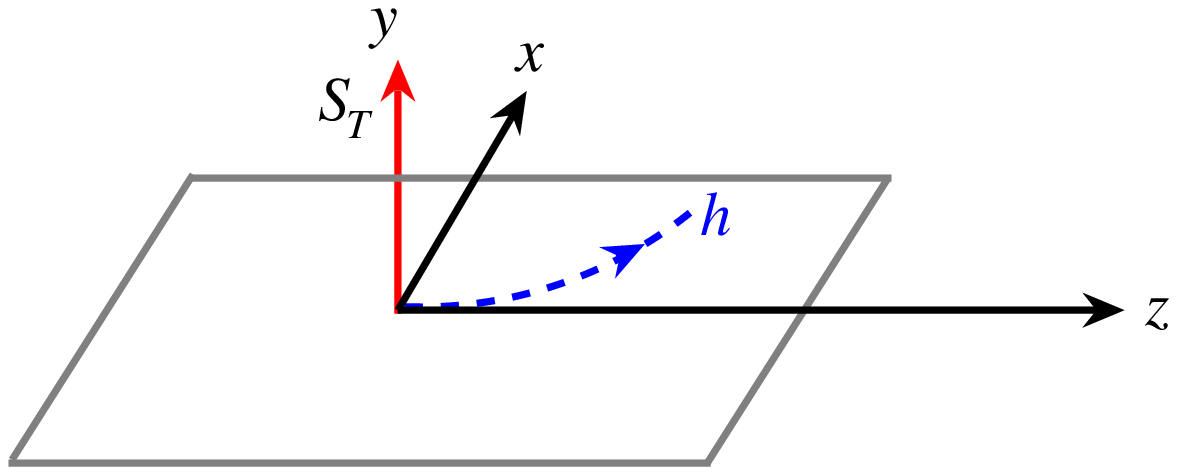, width=2.6in}
\caption{Illustration of the sign convention for $A_N$: positive
$A_N$ means that more hadrons are produced to the {\it left} 
of the beam direction when the beam's spin is vertically {\it
upward}.} \label{fig:exp} \eef 
We note at this point that there is an overall
sign error in \cite{oldpion} and consequently in \cite{newpion}, 
because in these papers the choice $\epsilon^{P_{h\perp}s_\perp n \bar{n}}>0$ 
was made (see Eq.~(73) of \cite{oldpion}, in contrast to Eq.~(\ref{epsilon}) 
above).

In the forward direction, $qg\to qg$ is the dominant partonic
scattering channel for inclusive single hadron production.
The corresponding hard-scattering functions are given by 
\cite{newpion} \ben
H^U_{qg\to qg}&=&\frac{N_c^2-1}{2N_c^2} \left[-\frac{\hat s}{\hat
u}-\frac{\hat u}{\hat s}\right]
\left[1-\frac{2N_c^2}{N_c^2-1}\frac{\hat s\hat u}{\hat t^2}\right]
\;\;\stackrel{|\hat t| \ll \hat s \sim |\hat u|}{\longrightarrow} \;\;
\left[\frac{2\hat s^2}{\hat t^2}\right],
\\
H^I_{qg\to qg}&=&\frac{1}{2(N_c^2-1)}
\left[-\frac{\hat s}{\hat u}-\frac{\hat u}{\hat s}\right]
\left[1-N_c^2\frac{\hat u^2}{\hat t^2} \right]
\;\;\stackrel{|\hat t| \ll \hat s \sim |\hat u|}{\longrightarrow} \;\;
\left[-\frac{N_c^2}{2(N_c^2-1)}\right]\left[\frac{2\hat s^2}{\hat t^2}
\right],
\\
H^F_{qg\to qg}&=&\frac{1}{2N_c^2(N_c^2-1)} \left[-\frac{\hat
s}{\hat u}-\frac{\hat u}{\hat s}\right] \left[1+2N_c^2\frac{\hat
s\hat u}{\hat t^2}\right] 
\;\;\stackrel{|\hat t| \ll \hat s \sim |\hat u|}{\longrightarrow} \;\;
\left[-\frac{1}{N_c^2-1}\right]\left[\frac{2\hat s^2}{\hat
t^2}\right] .
\een 
This shows that both $H^I_{qg\to qg}$ and $H^F_{qg\to qg}$ have 
opposite sign to that of the spin-averaged hard-scattering 
function $H^U_{qg\to qg}$. Furthermore it is clear that 
the SSA in $\pi^+$ production is mainly sensitive to $T_{u,
F}(x, x)$, while the one for $\pi^-$ production probes 
$T_{d, F}(x, x)$. Since \ben \frac{\epsilon^{P_{h\perp}s_\perp n
\bar{n}}}{\hat u}>0, \een 
we conclude from Eq.~(\ref{aftera}) that the observed positive SSAs for 
$\pi^+$ production indicates a {\it negative} $T_{u, F}(x, x)$, while the
observed negative asymmetry for $\pi^-$ production indicates a
{\it positive} $T_{d, F}(x, x)$, as shown by the solid curves in
Fig.~\ref{fig:sivers}.

To conclude this appendix, we demonstrate the apparent ``sign
mismatch'' again numerically, by evaluating the SSAs for inclusive
single hadron production using the ETQS functions indirectly
derived via Eq.~(\ref{rel}) 
from the quark Sivers functions in Eqs.~(\ref{old}) and
(\ref{new}). The results are shown in Fig.~\ref{fig:pion}.
As expected, the signs of the calculated SSAs are opposite to those
observed experimentally. 
\bef
\vspace*{5mm}
\psfig{file=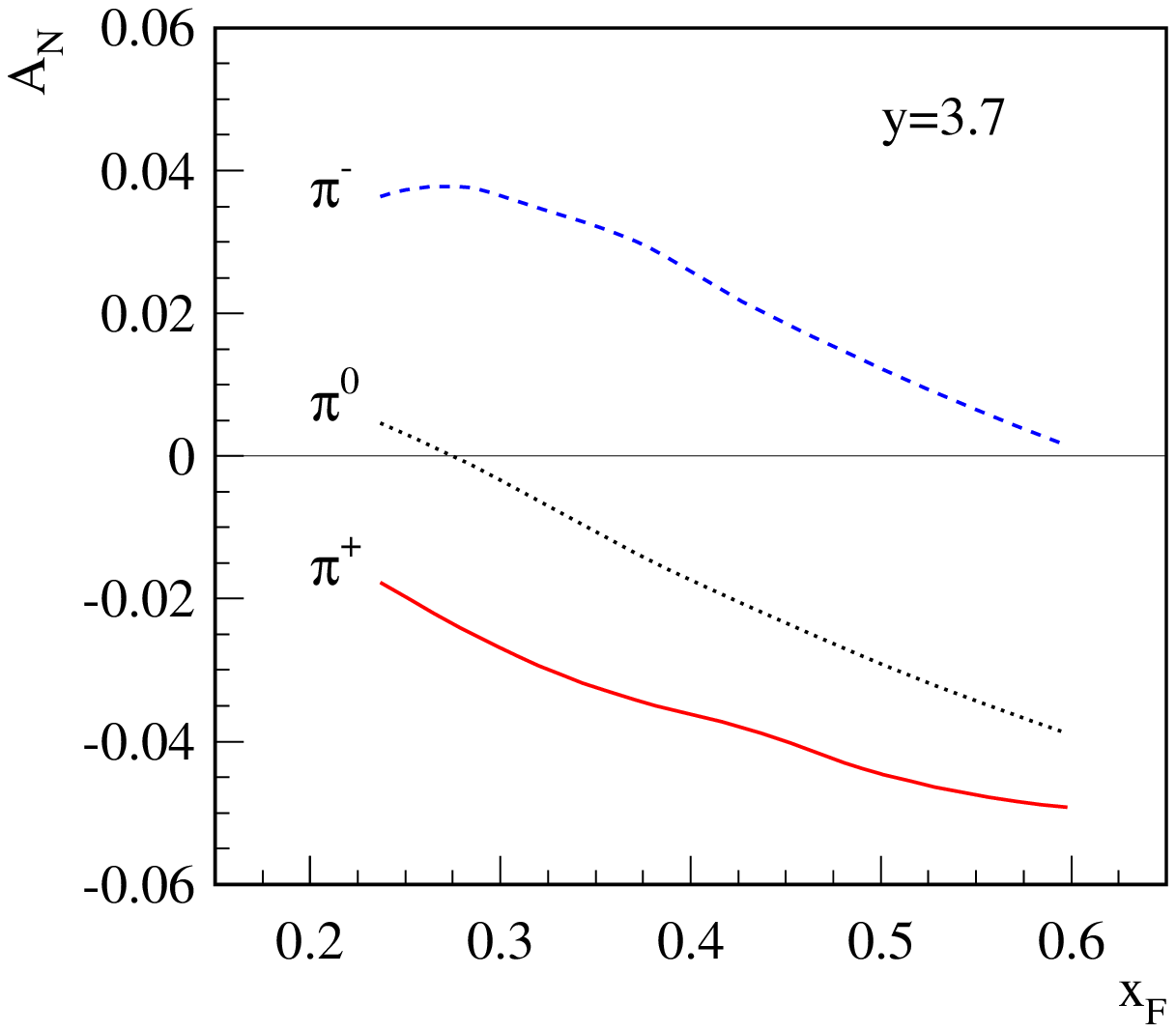, width=2.6in} \hskip 0.3in
\psfig{file=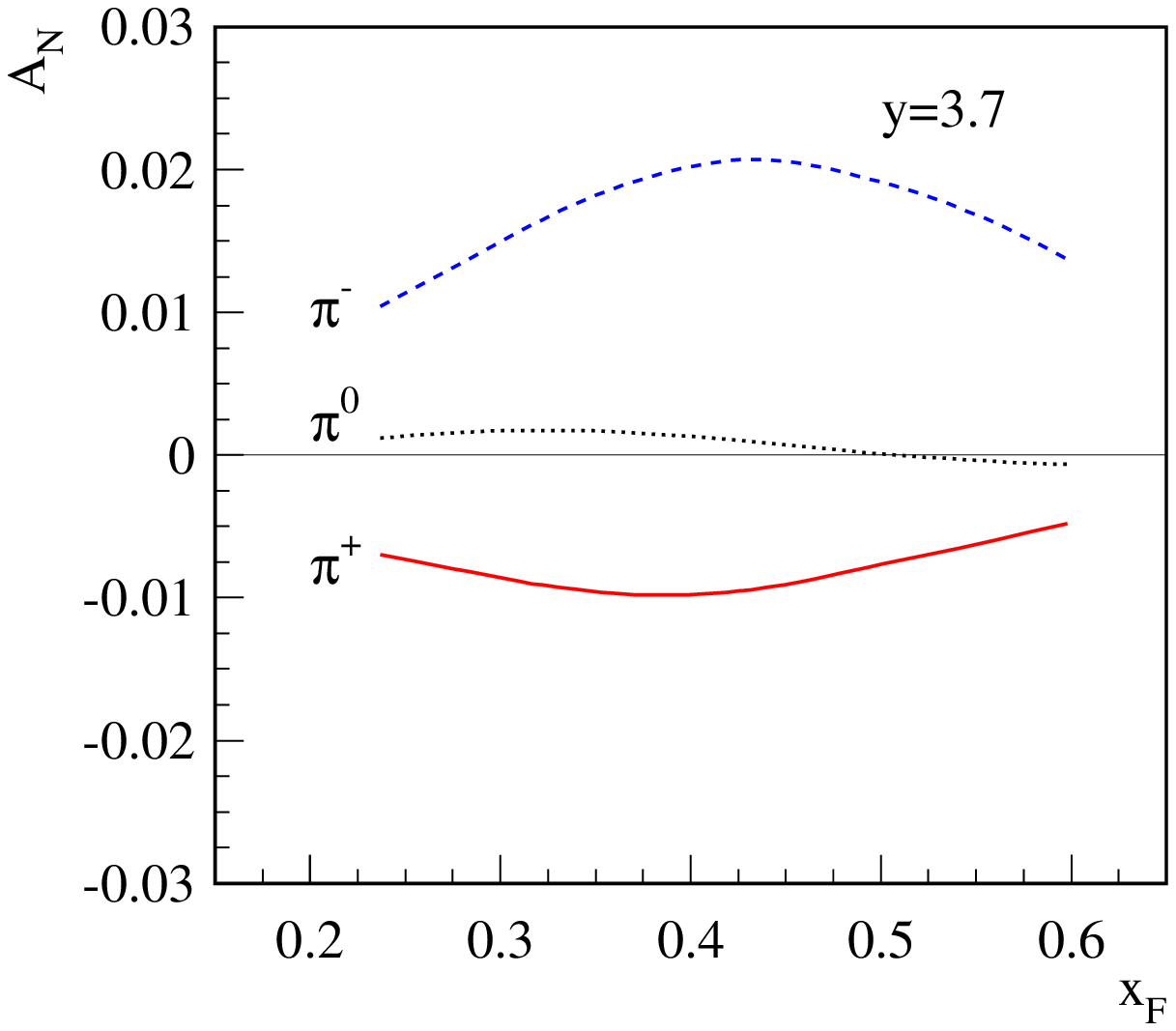, width=2.6in} \caption{The SSA, $A_N$,
for inclusive single pion production in $p^\uparrow p\to\pi+X$ 
at $\sqrt{s}=200$ GeV, as a function of $x_F$ and at 
rapidity $y=3.7$, evaluated by using the old Sivers functions in
Eq.~(\ref{old}) (left), and the new Sivers functions in
Eq.~(\ref{new}) (right).} \label{fig:pion} \eef

\end{document}